\documentclass[twocolumn,amsmath,amssymb,aps,pra,superscriptaddress,longbibliography,floatfix]{revtex4-1}
\usepackage{physics}
\usepackage{units}
\usepackage{amsmath}
\usepackage{amsthm}
\usepackage{url}
\usepackage{natbib}
\usepackage{textcase}
\usepackage{amssymb}
\usepackage{graphicx}
\usepackage{bm}
\usepackage{float}
\usepackage{multirow,microtype,color,relsize,ulem}
\usepackage{subfigure}
\usepackage[breaklinks=true]{hyperref}
\usepackage[utf8]{inputenc}
\usepackage[english]{babel}
\hypersetup{colorlinks=true,linkcolor=blue,citecolor=blue}
\hypersetup{linktocpage}
\usepackage{CJK}
\usepackage{breakcites}
\usepackage{float}
\usepackage{siunitx}

\begin{document}
\title{Classical and quantum time crystals in a levitated nanoparticle without drive}

\author{Yi Huang}
\affiliation{School of Physics and Astronomy, University of Minnesota, Minneapolis, MN 55455, USA}
\affiliation{Department of Applied Physics, Xi'an Jiaotong University, Xi'an, Shaanxi 710049, China}

\author{Qihao Guo}
\affiliation{Department of Applied Physics, Xi'an Jiaotong University, Xi'an, Shaanxi 710049, China}

\author{Anda Xiong}
\affiliation{School of Physics and Astronomy, University of Birmingham, Birmingham, UK}
\author{Tongcang Li}
\affiliation{Department  of  Physics  and  Astronomy, Purdue  University,  West  Lafayette,  IN  47907,  USA}
\affiliation{School  of  Electrical  and  Computer  Engineering, Purdue  University,  West  Lafayette,  IN  47907,  USA}
\affiliation{Birck Nanotechnology Center,  Purdue University,  West Lafayette,  IN 47907,  USA}
\affiliation{Purdue Quantum Science and Engineering Institute, Purdue University, West Lafayette, Indiana 47907, USA}

\author{Zhang-qi Yin}\email{zqyin@bit.edu.cn}
\affiliation{Center of Quantum Technology Research, School of Physics, Beijing Institute of Technology, Beijing 100081, China}

\date{\today}

\begin{abstract}
    Time crystal is defined as a phase of matter spontaneously exhibiting a periodicity in time. 
    Previous studies focused on discrete quantum time crystals under periodic drive. 
    Here, we propose a time crystal model based on a levitated charged nanoparticle in a static magnetic field without drive. 
    Both the classical time crystal in thermal equilibrium and the quantum time crystal in the ground state can emerge in the spin rotational mode, under the strong magnetic field or the large charge-to-mass ratio limit. 
    Besides, for the first time, the \emph{time polycrystal} is defined and naturally appears in this model. 
    Our model paves a way for realizing time crystals in thermal equilibrium.
\end{abstract}
\maketitle

\section{Introduction}
Time crystal is a phase which spontaneously breaks time translational symmetry in the ground state \cite{sacha2017,khemani2019brief}. 
In 2012, Wilczek \textit{et al.} proposed two models for time crystals. 
One is quantum \cite{Wilczek2012} while the other is classical \cite{Shapere2012}. 
Later, Li \textit{et al.} proposed that both the quantum space-time crystal and time quasicrystal can be realized experimentally using trapped ions~\cite{li2012}. 
Quantum time crystals have been discussed a lot in the following years \cite{Bruno2013a,Wilczek2013,Bruno2013b,Li2012a,bruno2013,watanabe2015,Huang2018}. 
A no-go theorem was proved that, for many-body systems with short range coupling and finite volume, the quantum time crystal does not exist in thermal equilibrium \cite{watanabe2015}. 

The no-go theorem can be bypassed if one considers systems in non-equilibrium. 
In this way, the discrete time crystal was theoretically proposed \cite{Sacha2015,else2016,Khemani2016,yao2017} and experimentally verified \cite{zhang2017,Choi2017}. 
Later, both the discrete space-time crystal and the discrete time quasicrystal were realized \cite{Smits2018,Giergiel2018,Autti2018,Pizzi2019}. 
The discrete time crystal has also been discussed in topological quantum computation \cite{Bomantara2018}, cold atom \cite{Ho2017,Huang2018a}, etc. 
Recently, using long range coupling Hamiltonian or interacting gauge filed, the no-go theorem can also be bypassed, and the existence of quantum time crystals in the ground state has been proposed \cite{hberg2019,kozin2019}.  
However, these models are not only practically challenging, but also facing debates on its feasibility now \cite{Syrwid2020,Ohberg2020,Khemani2020,Kozin2020}. 

On the other hand, attentions on the classical time crystal are relatively low \cite{Bains2017,das2018,Feng2018,LI2020135156,Easso2019,dai2019}. 
The original classical time crystal model contains singular solution points \cite{Shapere2012}, which are difficult to be tested in experiments.
The correspondence with the original classical time crystal is mostly found in cosmology \cite{Bains2017,das2018,Feng2018,LI2020135156,Easso2019}.
Recently, Shapere and Wilczek showed that the ``Sisyphus dynamics'' could arise in the effective motion of a planar charged particle subjected to the magnetic field, and the classical time crystal Lagrangians emerges in the effective theory of their systems \cite{Shapere2019}. 
However, the amplitude of the Sisyphus dynamics depends on the external perturbations, and disappears in the ground state. 
Besides, asymmetric mass parameters are required in this model, which are difficult to realize.

Here, we propose a scheme to realize a time crystal based on a levitated charged nanoparticle placed in an uniform magnetic field, where two of the rotational modes of the nanoparticle are trapped, while the third one (spin) rotates freely. 
By eliminating the two trapped rotational (torsional) modes, we show the nonzero angular velocity in the effective theory for the third spin rotational mode in our model.
For the classical model in thermal equilibrium, the angular velocity changes sign occasionally due to thermal fluctuations, but the absolute value (speed) is fixed.
This phenomena is similar to a spatial polycrystal, where the order parameter breaks the space continuous translational symmetry but retains the rotational symmetry.
Here, by showing the nonzero average speed of the spin rotational mode with small fluctuations, we introduce the \textit{time polycrystal} in our model, where the time reversal symmetry (or the rotational symmetry in the time domain) remains but the time continuous translational symmetry is broken.
On the other hand, once the system is cooled down with nonzero magnetic flux near the quantum ground state, it breaks both the time reversal and the time continuous translational symmetry simultaneously, and thus coincides with the conventional definition of quantum time crystals~\cite{Wilczek2012}.
Furthermore, we find that the conditions required for the time crystal phase in our model should be experimentally realizable.

\section{Experimental setup and the classical model}
\begin{figure}[t]
    \centering
    \includegraphics[width=0.7 \linewidth]{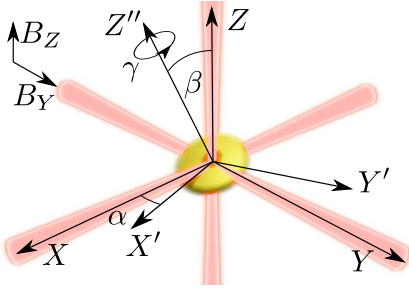}
    \caption{Schematic drawing of the experimental setup. A charged nanoparticle is trapped and levitated by external potential. All three translational degree of freedoms are fixed. The nanoparticle can freely rotate along the $Z''$ axis, but the other two rotational modes are doing small vibration. A strong uniform magnetic field $B_Y$ is applied along the $Y$ direction. Another weak magnetic field $B_Z \ll B_Y$ along the $Z$ direction can be used for achieving quantum time crystals.}
    \label{fig:exp}
\end{figure}

Before we present our theoretical model, let us introduce the possible experimental setup. 
As shown in Figure~\ref{fig:exp}, 
we consider a levitated charged insulating nanoparticle in an optical tweezer or an ion trap  \cite{Li2010,Yin2013,Fonseca2016,aranas2017thermometry}. 
Then the nanoparticle experiences both the force and torque in the trap until the mechanical equilibrium is reached where only the spin rotational mode is free. 
The three center of mass (c.m.) and other two rotational (torsional) modes are trapped \cite{Hoang2016,Ahn2018,Reimann2018,Ahn2020,Xiao2018,Bang2020}. 
In the following, we ignore the coupling between c.m. and rotations, and only consider the rotations and their own coupling~\footnote{This statement is justified for two reasons.
First, c.m. do not couple with rotations if we use harmonic traps.
Second, the nonlinear coupling between c.m. and rotations in non-harmonic traps (e.g. Gaussian traps in optical tweezers) is much smaller than the coupling among rotations themselves at low temperature, as shown in Ref.~\onlinecite{Xiao2018}}.
Next, the nanoparticle is pierced through by a strong uniform magnetic field.
Since the nanoparticle is charged, its rotation generates a magnetic moment coupling to the external magnetic field, which mixes the two torsional modes and the remaining spin rotation. 

Now let us examine the classical model.
The Lagrangian of a particle with the mass $m$ and the charge $Q$ inside the electromagnetic field is given by
\begin{equation}
	L = \frac{1}{2} m v^2 + Q \vb{v}\vdot \vb{A} - Q \Phi,
\end{equation}
where $\vb{A}$ is the vector potential and $\Phi$ is the electric potential. If the magnetic field is uniform, using the symmetric gauge $\vb{A} = \vb{B} \cross \vb{r}/2$, we arrive at the Lagrangian for a charged rigid body around its center of mass by integrating over the body volume
\begin{equation}\label{eq:L1}
	\begin{split}
		L &= \int \dd[3]{r} \left[\frac{1}{2}\rho_m(\vb{r}) (\vb*{\omega} \cross \vb{r})^2 \right.\\
		  &+ \left.\frac{1}{2}\rho_e(\vb{r}) (\vb*{\omega} \cross \vb{r}) \vdot (\vb{B} \cross \vb{r}) - \rho_e(\vb{r}) \Phi(\vb{r})\right].
	\end{split}
\end{equation}
After a simple algebra we arrive at 
\begin{equation}\label{eq:original_L}
	L = \frac{1}{2} m_{ij} \omega_i \omega_j + \frac{1}{2} e_{ij} B_i \omega_j - U(\vu{r}),
\end{equation}
where $U(\vu{r})$ is the trapping potential, $\omega_i$ and $B_i$ are the $i$th component of the angular velocity and the magnetic field respectively, $m_{ij}$ and $e_{ij}$ are tensors defined as $m_{ij} = \int \dd[3]{r} \rho_m(\vb{r}) (r^2 - r_ir_j)\qc$ and $e_{ij} = \int \dd[3]{r} \rho_e(\vb{r}) (r^2 - r_ir_j)$ with mass and charge density $\rho_m$ and $\rho_e$, and $i,j = 1,2,3$.
Using the $ZY'Z''$ Euler angles $(\alpha, \beta, \gamma)$, the Lagrangian can be written as
\begin{widetext}
\begin{align}\label{eq:original_L1}
	L &= \frac{1}{2} m_1 (\dot{\alpha}^2 \sin^2{\beta} + \dot{\beta}^2) + \frac{1}{2} m_3(\dot{\alpha} \cos{\beta} + \dot{\gamma})^2 + \frac{B_X}{4} [(e_3-e_1) \dot{\alpha} \cos{\alpha}\sin(2\beta) - 2e_1 \dot{\beta}\sin{\alpha} + 2e_3 \dot{\gamma} \cos{\alpha} \sin{\beta}] \nonumber\\
	  &+ \frac{B_Y}{4} [(e_3-e_1) \dot{\alpha} \sin{\alpha}\sin(2\beta) + 2e_1 \dot{\beta}\cos{\alpha} + 2e_3 \dot{\gamma} \sin{\alpha} \sin{\beta}] + \frac{B_Z}{2} [(e_1 \sin^2{\beta} + e_3 \cos^2{\beta}) \dot{\alpha} + e_3 \dot{\gamma} \cos{\beta}] -U(\alpha, \beta, \gamma).
\end{align}
\end{widetext}

If the potential $U(\alpha, \beta, \gamma)$ has equilibrium position $\alpha = \beta = 0$ but no constraint in $\gamma$, then $U\simeq u_1 \beta^2/2 + u_2 \alpha^2/2$. Suppose the magnetic field is $\vb{B} = (0,B_Y,B_Z)$, and $B_Z \ll B_Y$. Expand Eq.~\eqref{eq:original_L1} and keep only the leading terms of small angles, we arrive at 
\begin{align}\label{eq:leading}
	L &= \frac{1}{2}m_1 \dot{\beta}^2 + \frac{1}{2}m_3 (\dot{\alpha} + \dot{\gamma})^2 + \frac{B_Z e_3}{2} (\dot{\alpha} + \dot{\gamma})\nonumber \\
	  &+ \frac{B_Y}{2} \qty[e_3 \alpha \beta (\dot{\alpha} + \dot{\gamma}) + e_1\dv{t}\qty(\beta - \frac{1}{2}\alpha^2\beta)]\nonumber \\
	  &- u_1 \beta^2/2 - u_2 \alpha^2/2.
\end{align}
In the following we drop the gauge term (the total time derivative) in the squared bracket of second line of Eq.~\eqref{eq:leading}, because it is localized in space and contributes to neither the classical nor the quantum dynamics. 
As a result, we end up with the effective Lagrangian
\begin{equation}\label{eq:L2} 
	L = \frac{1}{2} m_1 \dot{x}^2 + \frac{1}{2} m_3 \dot{z}^2 + b xy\dot{z} + \phi\dot{z} - \frac{1}{2} u_1 x^2 - \frac{1}{2} u_2 y^2,
\end{equation}
where variables change is performed $x=\beta$, $y=\alpha$, $z=\alpha+\gamma$, and the notations $b \equiv B_Y e_3 / 2$ and $\phi \equiv B_Z e_3 / 2$ are used. 
The equations of motion are
\begin{align}
  m_1 \ddot x &= b y \dot z - u_1 x, \label{eq:eom_x}\\
  0 &= b x \dot z - u_2 y, \label{eq:eom_y}\\
  m_3 \ddot z &= - b \dv{(xy)}{t}. \label{eq:eom_z}
\end{align}
If we take limit $m_{1,3} \to 0$ in Eq.~\eqref{eq:original_L1}, the first two terms related to kinetic energy can be neglected. Substitution of Eq.~\eqref{eq:eom_y} into Eq.~\eqref{eq:eom_x} and \eqref{eq:eom_z} leads to 
\begin{gather}\label{eq:eom1}
	x(\dot{z}^2 - u_1 u_2 / b^2) = 0\\
	\dv{(x^2\dot{z})}{t} = 0.\label{eq:eom2}
\end{gather}
The energy of the system imposed by $m_{1,3} \to 0$ are given by
\begin{equation}
    E = \frac{1}{2}x^2(u_1 + b^2 \dot{z}^2/u_2).
\end{equation}
Therefore, the total energy can be arbitrarily close to the ground state energy by choosing the initial condition $x(t=0)$ small enough.
Forced by Eq.~\eqref{eq:eom1} and \eqref{eq:eom2}, the system rotates with a constant angular speed $\abs{\dot{z}} = \sqrt{u_1 u_2} / b$, as long as $x \neq 0$.
This effective dynamics reminds us the classical time crystal proposed by Shapere and Wilczek \citep{Shapere2012,Shapere2019}.
However, the small moment of inertia $m_{1,3}$ regularizes the pathological property of the classical time crystal Lagrangian, so we expect to see that the system rotates in a constant velocity near the ground state with an exitation energy proportional to $m_{1,3}$. 
As $m_{1,3} \to 0$, the system behaves as a classical time crystal. 

The correspondence between Eq.~\eqref{eq:L2} within limits $m_{1,3} \to 0$ and the classical time crystal encourages us to study the dynamics of Eq.~\eqref{eq:L2} when $m_{1,3}$ is small but finite.
Notice that the phase space of the system defined by Eq.~\eqref{eq:L2} has four instead of six DOFs.
Treating Eq.~\eqref{eq:eom_y} as a constraint and eliminating $y$ from Eq.~\eqref{eq:L2}, we arrive at the Lagrangian and the Hamiltonian (by doing Legendre transformation) 
\begin{gather}
	L =\frac{1}{2} m_1 \dot{x}^2 + \frac{1}{2} (m_3 + b^2x^2/u_2) \dot{z}^2 + \phi \dot{z} - \frac{1}{2} u_1 x^2,\label{eq:L}\\
	H = \frac{p_x^2}{2 m_1} + \frac{(p_z - \phi)^2}{2(m_3 + b^2 x^2 / u_2)} + \frac{1}{2}u_1 x^2, \label{eq:H_c}
\end{gather}

Notice the total time derivative term $\phi \dot{z}$ in Eq. \eqref{eq:H_c} has no effect in classical mechanics, so we ignore this term in the classical model. 
However, in quantum mechanics, this gauge term $\phi \dot{z}$ accumulates a geometric phase in the wavefunction, since $z$ can rotate a full cycle. 
This phase changes the quantum spectrum and leads to nonzero angular velocity in the ground state, i.e. a quantum analogy of the time crystal, as shown below.

\begin{figure}[t]
	\centering
	\includegraphics[width=\linewidth]{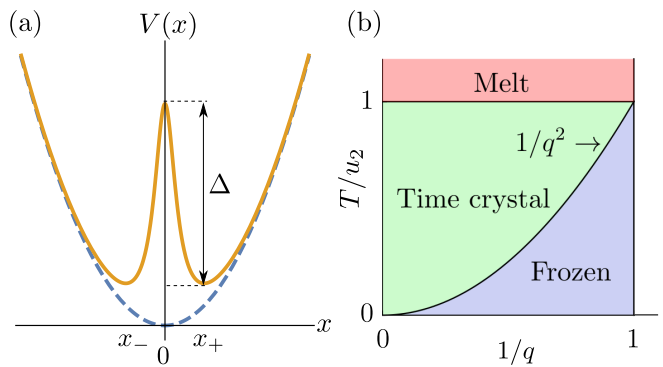}
	\caption{(a) Schematic picture of effective potential energy $V(x)$. If $\abs{l}<l_c$, $V(x)$ behaves as the blue curve (dashed); if $\abs{l}>l_c$, $V(x)$ behaves as the orange curve (solid), and there is a potential barrier $\Delta$ in the middle with three energy extrema $x = 0, x_{\pm}$. The latter has a Mexican hat shape, which implies spontaneously symmetry breaking. (b) Schematic phase diagram of the classical time crystal. Time crystal phase exists when $q>1$ and $T<u_2$ shaded as green. For $T > u_2$, the time crystal is ``melt'' (shaded as red); 
	while for $T < u_2/q^2$, the time crystal is ``frozen'' (shaded as blue).
	}
	\label{fig:schm}
\end{figure}

Let's return to the classical model.
Since $z$ is cyclic, $p_z = l$ is a constant. 
For a given $l$, the effective potential energy reads
\begin{equation}\label{eq:potential}
	V(x) = \frac{l^2}{2(m_3 + b^2 x^2 / u_2)} + \frac{1}{2}u_1 x^2,
\end{equation}
where the first term is similar to the centrifugal potential except that $V(x=0)$ is finite, while the conventional centrifugal barrier diverges as $x^{-2}$. 
There exists a critical angular momentum $l_c = m_3 \sqrt{u_1 u_2}/b$. 
If $0<\abs{l}<l_c$, there is one single equilibrium point $x=0$ which is stable.
The nontrivial case is $\abs{l} > l_c$, in which we have three extrema: $x_{\pm} = \pm [l_c(\abs{l}-l_c)/m_3 u_1]^{1/2}$ and $x_0=0$.
$x_{\pm}$ are stable with corresponding minima $V(x_\pm) =[l^2 -(\abs{l} - l_c)^2]/2m_3$, while $x_0$ is unstable with local maximum $V(0)=l^2/2m_3$. 
This creates a potential barrier $\Delta = (\abs{l}-l_c)^2/2m_3$ at the center.
The different potential profiles are shown in Figure~\ref{fig:schm} (a).
If we choose $x(t=0)=x_{\pm}$, then $\dot{z} = \si{sign}(l) \sqrt{u_1 u_2}/b$ is independent on $|l|$. 
Below we show in the numerical simulations that even if $x$ is doing small oscillation around $x_{\pm}$, the angular velocity averaged over a period is a constant $\overline{\dot{z}} = \si{sign}(l) \sqrt{u_1 u_2}/b$.
Since this property of $\overline{\dot{z}}$ is quite robust when $\abs{l}>l_c$, in the following we consider $\overline{\dot{z}}$ as an order parameter.
If $\overline{\dot{z}}= \si{sign}(l) \sqrt{u_1 u_2}/b$, we say that the system is in the (classical) time crystal phase.

For convenience, we define the quality of the time crystal phase as $q \equiv  b/\sqrt{m_3 u_{2}} = b/m_3 \omega_2$, where $\omega_{1,2} = \sqrt{u_{1,2}/m_3}$.
In order to get some intuitions of how the trajectory $x(t)$ looks like in the time crystal phase, one may consider the oscillation around $x_{\pm}$ and take limit $m_3 u_2 / b^2 \ll x^2$, or equivalently $q^2 x^2 \gg 1$, then the effective potential is replaced by $V(x) = l^2 u_2/2b^2 x^2 + u_1 x^2/2$.
This procedure is justified if the oscillation amplitude is small and $q^2 x_{\pm}^2 \gg 1$. 
Notice that since $q^2 x_{\pm}^2 = (\abs{l} - l_c)/l_c$, this limit $\abs{l}/l_c \gg 1$ is easily fulfilled within the time crystal phase.
The trajectory $x(t)$ can be solved analytically as $x(t)=\tfrac{1}{\sqrt{u_1}} \qty[E+\sqrt{E^2 - V^2(x_{\pm})} \sin(2\omega_1(t-t_0))]^{1/2}$, where $V(x_{\pm}) = \abs{l} \sqrt{u_1 u_2}/b$ are minimum of the double wells, $E > V(x_{\pm})$ is the total energy of the system, $t_0$ is a constant depends on the initial value $x(0)$. 
Thus $x(t)$ oscillates around $x_{\pm}$ with frequency $2\omega_1$, and in Appendix~\ref{app:velocity} we show the average angular velocity in one period is the same constant $\overline{\dot{z}} = \si{sign}(l) \sqrt{u_1 u_2}/b$.

The connection between our model to a classical time crystal can be seen as follows.
First, in the limit of $m_{1,3} \to 0$ (or $q \gg 1$), Eq.~\eqref{eq:L} reduces to exactly the same form as the classical time crystal Lagrangian in Ref.~\onlinecite{Shapere2012}.
Second, any finite but small $m_{1,3}$ will regulate the singularity of the classical time crystal Lagrangian, but the feature of a time crystal remains.
Although the true ground state of this system is at rest, small perturbations can drive the system into the time crystal phase.
Namely, if the initial condition satisfies $\abs{l} \gg l_c$, which can be fulfilled by arbitrarily small perturbation in the limit $q \to \infty$, then $x=0$ is no longer stable and the system will approach to $x_{\pm} \neq 0$, and as a result $\dot{z} =\si{sign}(l) \sqrt{u_1 u_2}/b$.
In Section~\ref{sec:numerical}, numerical simulations show that even if the initial perturbation is big, a small damping will force the system stay close to $x_{\pm} \neq 0$, with nontrivial angular velocity $\dot{z} = \si{sign}(l)\sqrt{u_1 u_2}/b$, until the energy is dissipated completely and the motion ceased. (See Figure~\ref{fig:damp}). 
In an imperfect vacuum, the lifetime of this rotation is inversely proportional to the pressure of the residual gas, and could be larger than $6\times 10^4$ s in experiments \cite{Rider2018}. 

Next we show a schematic picture of the phase diagram in the plane $(T, q)$ in Figure~\ref{fig:schm}(b).
The thermal fluctuations perturb the system with energy $E \sim T$ (here $k_B = 1$). 
To observe a time crystal phase, the temperature should be such that $T < u_2$, otherwise our small angle approximation fails and the Lagrangian Eq.~\eqref{eq:L} is invalid.
Therefore we say the time crystal is ``melt'' at $T > u_2$.
On the other hand, if $T$ is too small, there is no enough energy to reach the threshold angular momentum $l_c$.
(The effective potential is not double-well when $\abs{l}\le l_c$ and $\overline{\dot{z}} \neq \si{sign}(l) \sqrt{u_1 u_2}/b$). 
Thus, the velocity $\ev{\dot{z}} < \sqrt{u_1 u_2}/b$ and  will depend on the initial conditions of the system.
In this sense, we say the time crystal is ``frozen''.
Notice that the minimum energy to reach $l_c$ is $l_c^2/2m_3$, by equating $l_c^2/2m_3$ and $T$ we obtain the boundary between the time crystal phase and the frozen phase is $T = u_2/q^2$.
Therefore, the temperature range for the time crystal phase is $u_2/q^2 < T < u_2$, which exists only when $q>1$.
The higher the $q$, the larger the range of temperature in the time crystal phase, and hence the higher the quality of the time crystal.
In other words, the time crystal phase is only sensitive to a single dimensionless parameter, the quality $q$.

Below we will show the system in thermal equilibrium rotates with a specific speed but changes its rotational direction by thermal fluctuations. 
In this way, this thermal equilibrium state breaks the time translational symmetry and is doubly degenerate.
Thus, we call this state a classical time polycrystal.
As the time crystal is defined as a matter phase breaking the time translational symmetry, which presents wherever the absolute value $\abs{\dot{z}} \neq 0$ is fixed.
The further constraint that $\dot{z} \neq 0$ corresponds to the breaking of the time reversal symmetry, which is a sufficient but not necessary condition for a time crystal.
Since our system has time reversal symmetry, $\ev{\dot{z}} = 0$ is guaranteed, where $\ev{\dots}$ denotes the ensemble average in thermal equilibrium.
Thus, we should look at the speed $\ev{\abs{\dot{z}}}$ and its relative fluctuation $\Delta \abs{\dot{z}}/\ev{\abs{\dot{z}}}$.
In Appendix~\ref{app:stat} we find that the relative fluctuation is smaller than 1 if $q^2 T < 5.3 u_2$.
This justifies the existence of the time polycrystal in the thermal equilibrium.

Let us explore the experimental feasibility of our classical time crystal model. 
Considering a hollow nanoparticle made by the hexagonal Boron nitride (h-BN) with radius $\sim 1\si{\mu m}$, thickness $\sim 10 \si{nm}$, mass density $\sim 2 \si{g/cm^3}$, and surface charge density $0.025\si{e/nm^2}$ trapped with torsional frequency $\sim 100\si{Hz}$ \cite{JIANG2015589,Goldwater2018}, we can make the quality $q \sim 10$ by applying magnetic field $\sim 5\si{T}$~\cite{Berry:1997,juchem:2017}. 
In order to observe the classical time crystal, the temperature should be $10 \si{mK} \lesssim T \lesssim 50  \si{mK} $ (See Appendix~\ref{app:exp} \ref{app:stat}), which is reachable by feedback cooling \cite{Li2011NatPh}. 

\section{\label{sec:numerical} Numerical simulations for the full classical model}
Here we provide the form of dimensionless Lagrangian and Hamiltonian and the corresponding results of numerical simulations for the full classical model.
Generally speaking, each pair $m_{1,3}$ $e_{1,3}$, and $u_{1,2}$ are different only by some geometrical factors $\mu \equiv m_1/m_3$, $\nu \equiv e_1/e_3$, and $\eta \equiv \sqrt{u_2/u_1}$ depending on the shape of the nanoparticle. 
For a nano-ellipsoid or nano-dumbbell or other shapes of nanoparticles commonly used in levitated experiments, $\mu, \nu, \eta \sim 1$.
In order to proceed numerical simulations, we choose the units of energy and time as $E_0 = u_2 = m_3 \omega_2^2$ and $\tau_0 = 1/\omega_2$ respectively. 
Without lost of generality, we use $B_X = 0$ and the trapping potential
\begin{equation}
	U(\alpha, \beta) = -u_1 \cos{\beta} - u_2 \cos{\alpha},
\end{equation}
such that Eq.~\eqref{eq:original_L1} reduces to the following
\begin{align}\label{eq:dimless_Lc}
	L &= \frac{1}{2} \mu (\dot{\alpha}^2 \sin^2{\beta} + \dot{\beta}^2) + \frac{1}{2} (\dot{\alpha} \cos{\beta} + \dot{\gamma})^2 \nonumber\\
	  &+ \frac{q}{2} [(1-\nu) \dot{\alpha} \sin{\alpha}\sin(2\beta) + 2\nu \dot{\beta}\cos{\alpha} + 2 \dot{\gamma} \sin{\alpha} \sin{\beta}] \nonumber\\
	  &+ aq [(\nu \sin^2{\beta} + \cos^2{\beta}) \dot{\alpha} + \dot{\gamma} \cos{\beta}] + \eta^2 \cos{\beta} + \cos{\alpha},
\end{align}
where $a = B_Z/B_Y$.
Similarly, devide Eq.~\eqref{eq:L2} by $E_0$, we have 
\begin{equation}
	L = \frac{1}{2} \mu \dot{x}^2 + \frac{1}{2} \dot{z}^2 + q xy \dot{z} + aq \dot{z} - \frac{1}{2} \eta^2 x^2 - \frac{1}{2} y^2,
\end{equation}
and Eqs.~\eqref{eq:L} and \eqref{eq:H_c} reduce as 
\begin{gather}
	L = \frac{1}{2} \mu \dot{x}^2 + \frac{1}{2} (1 + q^2 x^2) \dot{z}^2 + aq\dot{z} - \frac{1}{2} \eta^2 x^2, \\
	H = \frac{p_x^2}{2\mu} + \frac{(p_z - aq)^2}{2(1+q^2x^2)} + \frac{1}{2} \eta^2 x^2. \label{eq:h_dimless}
\end{gather}
Since generally $\mu, \nu, \eta \sim 1$, it is clear that the system is sensitive to only one dimensionless parameter, i.e. the quality of time crystal phase $q$.
In order to get time crystal behavior in numerical simulations, we choose $q \gg 1$.
The value of $a$ does not affect classical mechanics, but will play a role in quantum mechanics.

\begin{figure}[t]
	\centering
	\includegraphics[width=\linewidth]{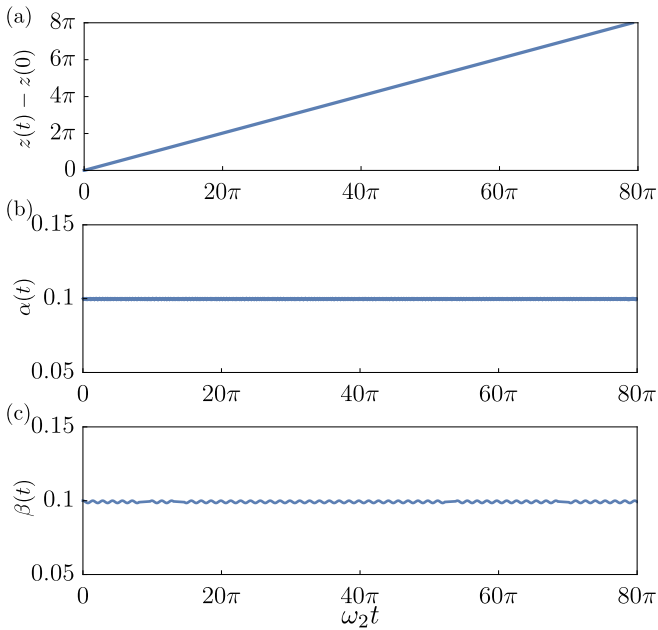}
	\caption{$z(t) = \alpha(t) + \gamma(t)$, $\alpha(t)$, and $\beta(t)$ as a function of time. The parameters are $\mu = \nu = \eta = 1$ and $q=10$, damping coefficient $\delta = 0$. Initial conditions are $\alpha(0) = \beta(0) = 0.1$, $\dot{\alpha}(0) = \dot{\beta}(0) = \gamma(0) = 0$, and $\dot{\gamma}(0) = 1/q$, where the unit of velocity is $\omega_2$. (a) $z(t) = \alpha(t) + \gamma(t)$. The angular velocity is $\dot{z} = \omega_2/q = \sqrt{u_1 u_2}/b$ as expected. (b) $\alpha(t)$. (c) $\beta(t)$. Both $\alpha$ and $\beta$ are essentially constants, which corresponds to the minimum of the double well potential.}
	\label{fig:z}
\end{figure}
\begin{figure}[t]
	\centering
	\includegraphics[width=\linewidth]{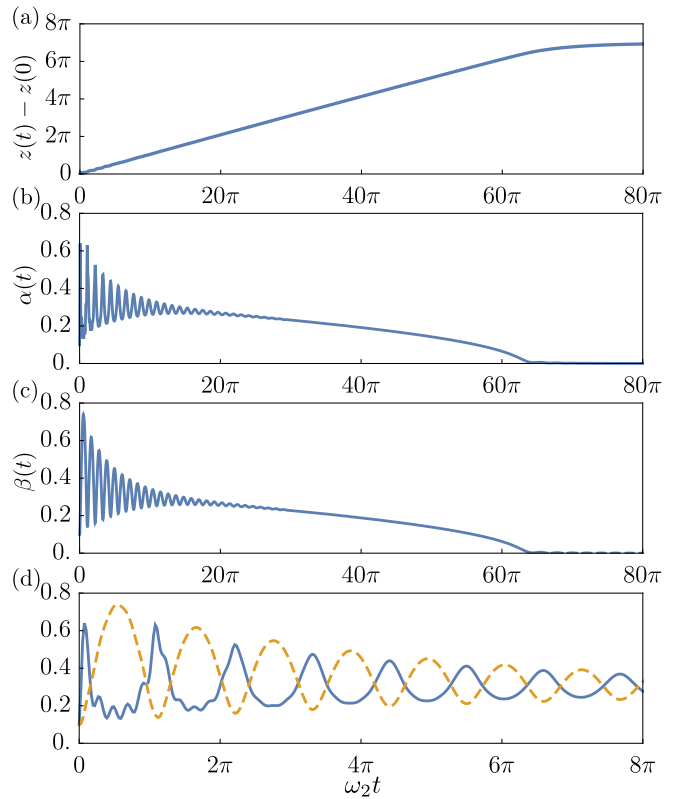}
	\caption{Damped motion. The parameters are the same as Figure~\ref{fig:z}. We assume the damping coefficient $\gamma_d = 0.05$. Initial conditions are $\alpha(0) = 0.1$, $\beta(0) = 0.1$, $\dot{\alpha}(0) = \dot{\beta}(0) = \gamma(0) = 0$, and $\dot{\gamma}(0) = 10/q$. (a) Although $z(t) = \alpha(t) + \gamma(t)$ first deviates from $z = \omega_2 t/q$ in the beginning, the damping forces $\dot{z}$ stabilized at $\dot{z} = \omega_2/q = \sqrt{u_1 u_2}/b$, until the energy is not enough to support the rotation and it suddenly ceases. (b) $\alpha(t)$. (c) $\beta(t)$. The oscillations get damped until $\alpha$ and $\beta$ become equal, and then slowly decay to zero. (d) Comparison between $\alpha$ (solid line) and $\beta$ (dashed line) in the interval $t \in (0,8\pi)$. }
	\label{fig:damp}
\end{figure}

Next we discuss results of numerical simulations of the classical model Eq.~\eqref{eq:dimless_Lc}.
In Figure~\ref{fig:z}, we show the motion $z(t) = \alpha(t) + \gamma(t)$ with initial velocity $\dot{\gamma}(0) = 1/q$.
It is clear that $\dot{z} = 1/q$ though there are small initial perturbations in $\alpha(0) = 0.1$ and $\beta(0) = 0.1$.
Further simulation shows that $\dot{z} = 1/q$ is quite robust even if we double the initial velocity as $\dot{\gamma}(0) = 2/q$ and keep perturbations in $\alpha$ and $\beta$ small.
We find $\dot{z}$ start deviates from $1/q$ as $\dot{\gamma}(0) \gtrsim 5/q$ while keeping $\alpha(0)$ and $\beta(0)$ as 0.1.
However, if we add small damping term linear in velocity such that equations of motion are given by 
\begin{equation}
    \dv{t} \pdv{L}{\dot{x}_i} = \pdv{L}{x_i} - \gamma_d \dot{x}_i,
\end{equation}
where $x_i = x, y, z$, and $\gamma_d$ is the decay coefficient. We find that $\dot{z}$ stabilizes again at $1/q$ even if we make big perturbation to the initial conditions, as shown in Figure~\ref{fig:damp}.
The reason that $\dot{z} = 1/q$ stabilizes under damping is that the system slowly falls into one of the minimum of the double-well potential shown in Figure~\ref{fig:schm}(a).
This state at the minimum of the double well is exactly in the classical time crystal phase.

\section{quantum model}
In this section, we study whether the time crystal phase exists in the quantum analog of our model. 
Before we proceed to the quantum model, it is heuristic and pedagogical to consider the semi-classical approach by applying Bohr-Sommerfeld quantization condition $\oint p_x \dd{x} = (n + 1/2)h$, where $n \in \mathbb{Z}$ and $p_x = \sqrt{2m(E-V(x))}$ is the momentum. 
Again we take limit $m_3 \to 0$ to get analytical results. 
The spectrum is $E_n = V(x_{\pm}) + \hbar\omega_1(2n+1)$, which looks like a simple harmonic oscillator (SHO) with frequency $2\omega_1$, and is consistent with the frequency of the classical trajectory.

Now we are well-prepared to solve the quantum model. 
The Schr\"{o}dinger equation $H \Psi = E \Psi$ can be solved by separation of variables $\Psi(x,z) = \chi(x)\zeta(z)$, which leads to two ordinary differential equations 
\begin{gather}
	-\frac{\hbar^2 \chi''}{2m_1} +\qty(\frac{\hbar^2 \sigma(l)}{2m_1(x^2 + \xi^2)} + \frac{1}{2}m_1 \omega_1^2 x^2) \chi = E \chi, \label{eq:qm_x}\\
	\hbar^2 \zeta'' = -l^2 \zeta,\label{eq:qm_z}
\end{gather}
where $\sigma = l^2 m_1 u_2/b^2\hbar^2$, $\xi^2 = m_3 u_2/b^2$. The boundary conditions are $\chi(\pm\infty) = 0$, $\zeta(z+2\pi) = \zeta(z)$ respectively. 
Eq.~\eqref{eq:qm_z} is trivial with solution $\zeta(z) = \mathrm{e}^{ilz/\hbar}$, where $l/\hbar \in \mathbb{Z}$. 
To solve Eq.~\eqref{eq:qm_x} analytically, we assume $\xi^2 \ll x^2$ which will be justified below. 
Following this assumption, the energy spectrum is determined by two quantum numbers $n$ and $l$~\cite{dong2007} 
\begin{equation}\label{eq:energy}
	E_{nl} = \qty(2n+1 + \sqrt{\sigma+1/4}) \hbar \omega_1. 
\end{equation}
We are interested in the expectation value of the angular velocity $\dot{z}$
\begin{equation}\label{eq:velocity}
	\ev{\dot{z}} = \frac{i}{\hbar} \ev{[H,z]} = \si{sign}(l) \frac{\sqrt{u_1 u_2} / b}{\sqrt{1+ (4\sigma)^{-1}}},
\end{equation}
which is true for all eigenstates $\ket{n l}$.
Notice that the mechanical momentum $\dot{z}$ is not the same as the canonical momentum $p_z = l$.
However, it is $\dot{z}$ that determines how fast the system rotates.
If $\sigma \gg 1$, then $\ev{\dot{z}} \to \si{sign}(l) \sqrt{u_1 u_2} / b$, which recovers the classical result; if $\sigma \ll 1$, then $\dot{z} \to 2l u_2 \sqrt{m_1 u_1} / b^2 \hbar$, which is linear in $l$. 
Furthermore, Eq.~\eqref{eq:velocity} allows us to evaluate the time correlation function $\ev{z(0)z(t)} = \ev{z(0)^2} + \ev{z(0)}\ev{\dot{z}(0)}t$, which linearly depends on time, and is true for any eigenstate. 
Since $z\in [0, 2\pi)$ is an angular variable, $\ev{z(0)z(t)}$ is actually periodic in $t$, with frequency $\propto \ev{\dot{z}(0)}$. 
\begin{figure}[t]
	\centering
	\includegraphics[width=\linewidth]{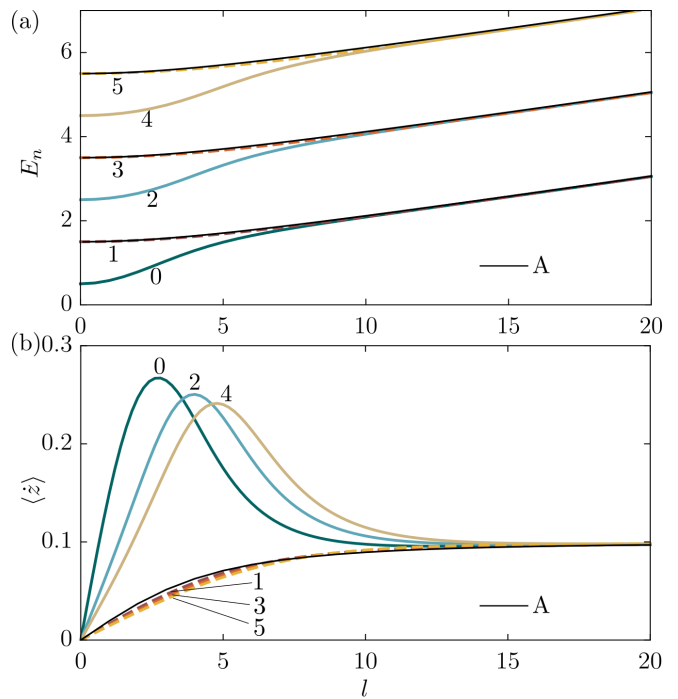}
	\caption{Spectrum $E_n$ and velocity $\ev{\dot{z}}$, presented in units such that $\hbar=\omega_1 = 1$, and $q = 10$. Numbers mark the curves with different quantum number $n$. Solid lines correspond to even $n$ (symmetric states) while dashed lines correspond to odd $n$ (antisymmetric states). The black curves marked by ``A'' correspond to Eqs.~(\ref{eq:energy},~\ref{eq:velocity}). (a) Spectrum $E_n$ as a function of angular momentum $l$. (b) $\ev{\dot{z}}$ as a function of $l$.}
	\label{fig:num_q}
\end{figure}

In Figure~\ref{fig:num_q}, we compare our analytical results Eq.~(\ref{eq:energy},~\ref{eq:velocity}) (shown as black curves marked by ``A'') with numerical simulations where $l\geq 0$ and $q=10$
\footnote{We find the qualitative behaviors are the same for all $q>1$; while for $q<1$, the contribution from the centrifugal potential is trivial (being a constant energy background) with $x$ dependence being suppressed.}.
The most important feature resides in the parity of wavefunctions.
For antisymmetric states with odd $n$, numerical results are practically the same as Eqs.~(\ref{eq:energy},~\ref{eq:velocity}); while for symmetric states with even $n$, numeric curves coincide with Eqs.~(\ref{eq:energy},~\ref{eq:velocity}) only when $l\gg q$ (in units shown in Figure~\ref{fig:num_q}).
This justifies that our assumption $\xi^2 \ll x^2$ is true for antisymmetric states independent on $l$, but true for symmetric states only for $l\gg q$.

Let us look at the energy spectrum shown in Figure~\ref{fig:num_q} (a). 
For small $l$, the spectrum is the same as the SHO $E_n = n+1/2$; while for large $l$ the spectrum approaches the SHO plus an inverse square potential, and each energy level is doubly degenerate.
It is even more interesting to look at $\ev{\dot{z}}$ as shown in Figure~\ref{fig:num_q} (b).
For symmetric states there is a hump at $l \sim q/2$, while the hump disappears for antisymmetric states.
Why is there a hump?
Recall that the operator $\dot{z}$ has the form $\sim p_z/(1+q^2x^2)$ maximized at $x=0$.
Since the operator $\dot{z} \propto p_z$, we expect $\ev{\dot{z}}$ grows linearly in $l$ at small $l\ll q$.
The antisymmetric states $\psi_A$ must vanish at $x=0$, and thus reduce $\ev{\dot{z}}{\psi_A}$; while symmetric states $\psi_S$ can be nonzero at $x=0$, and $\ev{\dot{z}}{\psi_S}$ gets enhanced.
This explains why the states with even $n$ rotates faster than the states with odd $n$.
However, the potential barrier starts to manifest itself and suppress $\psi_S(x=0)$ at $l\gtrsim q$, so the results for states with different parity converge at large $l$.
This qualitatively explains the hump appeared with odd $n$.
Numerical calculations show that the larger the $q$, the higher the hump, which demonstrates a sharp transition of $\dot{z}$ at $l\sim q/2$.

\begin{figure}[t]
	\centering
	\includegraphics[width=\linewidth]{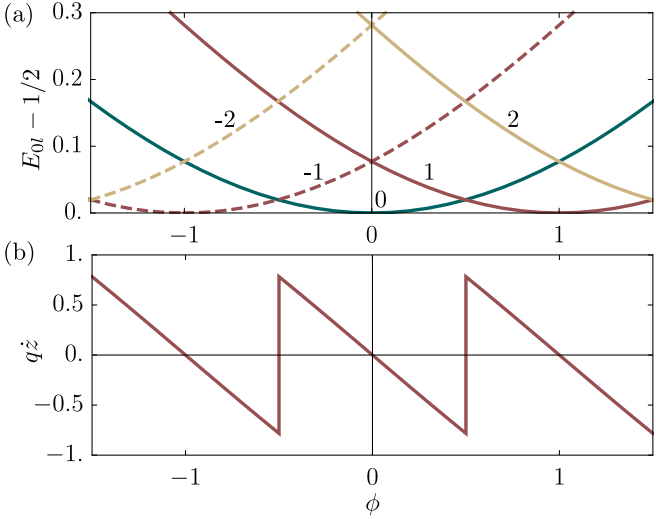}
	\caption{(a) Ground state energy $E_{0l} - 1/2$ as a function of flux $\phi$, where $\hbar=\omega_1 = 1$ and $q = 10$.  The quantum number $l$ is marked on each curve. (b) $\ev{\dot{z}}$, which is experimentally measurable angular velocity,  of the ground state as a function of magnetic flux $\phi$, where $\hbar=\omega_1 = 1$ and $q = 10$. We note that when $\ev{\dot{z}} \neq 0$, the time-translational symmetry is broken.}
	\label{fig:phi}
\end{figure}

For this system, the angular velocity vanishes for the ground state $n=l=0$. 
However, it is possible to get both $\ev{\dot{z}} \neq 0$ and periodic $\ev{z(0)z(t)}$ even for the ground state. 
This indicates the existence of quantum time crystal~\cite{watanabe2015}.
To present the quantum time crystal, we consider the consequence of $\phi \neq 0$.
It is well-known that this magnetic flux changes the spectrum, such that $\sigma \propto (l-\phi)^2$ in Eq.~\eqref{eq:energy}. 
The ground state energy traces on the bottom of the curves $E_{0l}(\phi)$ corresponding to different $l$ shown in Figure~\ref{fig:phi} (a).
Compare with $\dot{z}(\phi)$ shown in Figure~\ref{fig:phi} (b), we find $E_{0l}(\phi) = 0$ reaches its minimum while $\dot{z}(\phi)$ vanishes at $\phi \in \mathbb{Z}$. 
On the other hand, when $\phi$ is half integers, both $E_{0l}(\phi)$ and $\abs{\dot{z}(\phi)}$ get maximized while $\dot{z}(\phi)$ jumps between its minimum and maximum values.
More generally, as long as $\phi/\hbar \notin \mathbb{Z}$, $\sigma$ is non-vanishing even for the ground state, leading to nonzero $\ev{\dot{z}}$ given by Eq.~\eqref{eq:velocity}.
As a result, $\ev{\dot{z}}$ becomes a periodic function in $\phi$, as shown in Figure~\ref{fig:phi} (b).
Distinguished from the classical model, here the time reversal symmetry is firstly broken by the magnetic flux, leading to $\ev{\dot{z}} \neq 0$, and then the time translational symmetry is broken simultaneously.
In order to observe the quantum ground-state behavior experimentally, we require the temperature to be smaller than the excitation gap $\sim \hbar \omega_1 \simeq 1 \si{nK}$, which should be realizable by sympathetic cooling with BEC \cite{Ranjit2015}. 

\section{conclusion}
In conclusion, we have proposed a scheme to realize the time crystal in a levitated charged nanoparticle under a uniform static magnetic field. 
Both the classical time polycrystal in thermal equilibrium and the quantum time crystal in the ground state appear under the strong magnetic field or the large charge-to-mass ratio limit in our model.  
Thanks to the recent rapid developments of levitating, cooling, and manipulating the motion of the nanoparticle \cite{Li2010,Hoang2016,Ahn2018,Reimann2018,Ahn2020}, the conditions required for the time crystal phase in our model should be feasible in laboratory. 
Our study may stimulate the research on the space-time crystal and the time quasicrystal in other classical systems, e.g. the trapped ions crystals \cite{Mitchell1998,Li2017,Wang2019}. 
In future, it would be interesting to extend our time crystal model and study its spontaneous time transnational symmetry breaking in the framework of field theory. 


\appendix
\section{\label{app:velocity} Averaged angular velocity}
In this appendix, we calculate the averaged angular velocity in both the classical and quantum models.

\textit{1. Classical model.}
Using the trajectory $x(t)$ given in the main text, we can calculate the the angular velocity averaged in a period
\begin{align}
	\overline{\dot{z}} &= \frac{\omega_1}{\pi} \int_0^{\pi/\omega_1} \frac{\dd{t} l u_1 u_2}{b^2 \qty[E + \sqrt{E^2 - V(x_{\pm})^2} \sin(2\omega_1 t)]} \nonumber \\
					   &= \frac{\omega_1}{\pi} \frac{l u_1 u_2}{b^2} \frac{1}{2\omega_1} \frac{2\pi}{V(x_{\pm})} = \si{sign}(l) \frac{\sqrt{u_1 u_2}}{b}.			   
\end{align}

\textit{2. Quantum model.}
Here we prove Eq.~\eqref{eq:velocity} in the main text.  
The angular velocity is given by 
\begin{equation}
	\ev{\dot{z}} = \frac{i}{\hbar}\ev{[H, z]}{nl} = \frac{u_2 l}{b^2}\int_0^{\infty} \dd{x} \frac{\braket{nl}{x}\braket{x}{nl}}{x^2} ,
\end{equation}
The wave functions are 
\begin{equation}
	\braket{x}{nl} = \frac{(2n!)^{1/2} \rho^{s} \mathrm{e}^{-\rho/2}}{\Gamma(n+2s+1/2)^{1/2}}  L_n^{2s - 1/2}(\rho),
\end{equation}
where $\rho = x^2 m_1 \omega_1 / \hbar$ and $s = (1+\sqrt{1+4\sigma})/4$.
We will show that 
\begin{equation}
	\int_0^{\infty} \dd{x} \abs{\braket{x}{nl}}^2 /x^2 = \frac{m_1 \omega_1}{\hbar} \frac{2}{\sqrt{1+4\sigma}}.
\end{equation}
This is equivalent to prove the following mathematical identity
\begin{equation}
	\frac{n!}{\Gamma(n+\alpha+1)}\int_0^{\infty} \dd{x} x^{\alpha - 1} \mathrm{e}^{-x} L_n^{\alpha}(x)^2 = 1/\alpha.
\end{equation}

\begin{proof}
\begin{align*}
	&\int_0^{\infty} \dd{x}  x^{\alpha - 1} \mathrm{e}^{-x} L_n^{\alpha}(x)^2 \\
	&= \eval{\frac{1}{\alpha} x^{\alpha} \mathrm{e}^{-x} L_n^{\alpha}(x)^2}_0^{\infty} - \frac{1}{\alpha} \int x^{\alpha} \dd\qty(\mathrm{e}^{-x} L_n^{\alpha}(x)^2) \\
	&= \frac{1}{\alpha}\int_0^{\infty} \dd{x}  x^{\alpha} \mathrm{e}^{-x} (L_n^{\alpha}(x)^2 - 2 L_n^{\alpha} \dv*{L_n^{\alpha}}{x}).
\end{align*}
Notice that 
\begin{equation}
	\dv{L_n^{\alpha}}{x} = - L^{\alpha+1}_{n-1} = - \sum_{m=0}^{n-1} L_m^{\alpha},
\end{equation}
and the orthogonality of associated Laguerre polynomials
\begin{equation}
	\int_0^{\infty} \dd{x} x^{\alpha} \mathrm{e}^{-x} L_n^{\alpha}  L_m^{\alpha} = \frac{\Gamma(n+\alpha+1)}{n!} \delta_{mn},
\end{equation}
we arrive at 
\begin{equation}
	\int_0^{\infty} \dd{x}  x^{\alpha} \mathrm{e}^{-x} L_n^{\alpha} \dv*{L_n^{\alpha}}{x} = 0,
\end{equation}
and 
\begin{equation}
	\int_0^{\infty} \dd{x}  x^{\alpha - 1} \mathrm{e}^{-x} L_n^{\alpha}(x)^2 = \frac{\Gamma(n+\alpha+1)}{n!} \frac{1}{\alpha}.
\end{equation}

\end{proof} 

\section{\label{app:exp}Experimental applicability}
In this appendix, we estimate how large the quality $q$ could be in experiments, and the temperature range to observe the time crystal phase.
First we introduce the intrinsic frequency of the system $\omega_0 \equiv b/m_3 = B_Y e_3/2 m_3$, such that $q = \omega_0/\omega_2$. 
Since $e_3/m_3$ is essentially the charge-to-mass ratio of the particle, the larger the ratio, the higher the quality of the time crystal for a fixed magnetic field. 
In order to increase the ratio of charge-to-mass, consider the hollow nanoparticles, such that its mass is proportional to the surface area. 
On the other hand, the number of charges is also proportional to the surface area, so the charge-to-mass ratio doesn't depends of the radius of a hollow nanoparticle as long as its thickness is small compared to the radius.

Consider a nanoparticle made by the hexagonal Boron nitride (h-BN) with thickness $\sim 10\si{nm}$, mass density $\sim 2 \si{g/cm^3}$, and surface charge density $\sim 0.025\si{e/nm^2}$, such that its charge-to-mass ratio is $\sim 200 \si{C/kg}$. 
Notice that the average distance between nearest-neighbor charges is $\sim 6 \si{nm}$, which is much larger than the Bohr radius in h-BN $\sim 2 \si{\angstrom}$, which justifies our assumption that surface charges are classical point charges with long-range interaction in the main text.
If the magnetic field is $\sim 5\si{T}$, then the intrinsic frequency is $\sim 1\si{kHz}$.
Thus $\omega_2$ must be much smaller than $1\si{kHz}$ in order to have high quality of the time crystal.
For example, to make the quality $q \sim 10$, we require the oscillation frequency $\omega_2 \sim 100 \si{Hz}$. 
However, since $u_2 = m_3 \omega_2^2$ and the time crystal phase can only be observed for $u_2/q^2 < T < u_2$, the smaller the $\omega_2$, the more difficult to cool down the nanoparticle to the time crystal phase. 
To compensate the effect of small $\omega_2$, we want $m_3$ big enough such that $u_2$ is realizable in experiments. 
Notice that the moment of inertia $m_3$ is essentially the product of mass and radius square, we can control $m_3$ by increasing the radius of the nanoparticle and meanwhile fix its thickness.
For the nanoparticle we consider above, if we require $u_2 = 1\si{K}$, then the radius $\sim 1\si{\mu m}$.

As a result, remarkably, if we put a hollow particle with radius $\sim 1\si{\mu m}$ and thickness $\sim 10\si{nm}$ inside the optical or ion trap such that the twisting frequency is $\sim 100\si{Hz}$, then a classical time crystal phase should appear within the temperature $10 \si{m K} \lesssim T \lesssim 1\si{K}$ when we charge it and turn on a uniform magnetic field.
In the discussion above we ignore the thermal fluctuation of $\abs{\dot{z}}$.
Below we will show that the requirement of small fluctuation of $\abs{\dot{z}}$ in the time crystal phase should restrict the temperature range to $10 \si{m K} \lesssim T \lesssim 50 \si{m K}$.

\section{\label{app:stat}Statistical average of angular speed}
\begin{figure}[b]
	\centering
	\includegraphics[width=0.8\linewidth]{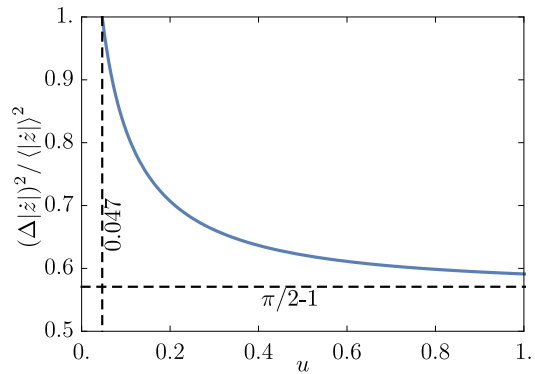}
	\caption{Relative fluctuation as a function of $u$.}
	\label{fig:fluc_c}
\end{figure}

Here we derive the upper bound of the temperature in the time crystal phase.
First we study the classical statistical average of angular speed $\abs{\dot{z}}$ from Eq.~\eqref{eq:h_dimless} in the main text with $\mu = \eta = 1$.
The partition funtion reads
\begin{equation}
	\mathcal{Z} = \tr \mathrm{e}^{-H/T} = 8\pi^{5/2} q T^2 U(-1/2,0,2u),
\end{equation}
where we define a scaling variable $u = (4q^2 T)^{-1}$ for convenience.
As $u \to 0$, $\mathcal{Z} \to 8\pi^2 q T^2$; while as $u \to \infty$, $\mathcal{Z} \to (2\pi)^{5/2} T^{3/2}$.
The absolute value of $\dot{z}$ reads
\begin{equation}
	\ev{\abs{\dot{z}}} =[q\sqrt{\pi} U(-1/2,0,2u)]^{-1}, 
\end{equation}
where $U$ is the confluent hypergeometric function, and $\ev{\abs{\dot{z}}} \to 1/q$ as $u \to 0$; while $\ev{\abs{\dot{z}}} \to \sqrt{2T/\pi}$ as $u \to \infty$.
Next we want to calculate the fluctuation of $\abs{\dot{z}}$
\begin{equation}
	(\Delta \abs{\dot{z}})^2 = \frac{\ev{\abs{\dot{z}}^2} - \ev{\abs{\dot{z}}}^2}{\ev{\abs{\dot{z}}}^2}.
\end{equation}
We start with the second moment 
\begin{equation}
	\ev{\abs{\dot{z}}^2} = \frac{\mathrm{e}^{u}K_0(u)}{2q^2 \sqrt{\pi} U(-1/2,0,2u)}.
\end{equation}
where $K_0$ is the zeroth-order second-typed Bessel function.
Therefore we have
\begin{equation}
	(\Delta \abs{\dot{z}})^2/\ev{\abs{\dot{z}}}^2 = \frac{\sqrt{\pi}}{2} \mathrm{e}^{u}K_0(u) U(-1/2,0,2u)-1.
\end{equation}
which only depends on the scaling variable $u$.
For $u>0.047$, we get that $\Delta \abs{\dot{z}} < \ev{\abs{\dot{z}}}$, and the lower bound of the relative fluctuation is $\pi/2 -1$ as shown in Figure~\ref{fig:fluc_c}.
The condition $u>0.047$ or equivalently $q^2 T < 5.3 u_2$ corresponds to $T < 53 \si{m K}$ if $q=10$ and $u_2 = 1\si{K}$. 
This justifies the upper bound of temperature range in the time crystal phase discussed in the last section and the main text.


%

\end{document}